\documentclass[conference]{IEEEtran}

\usepackage{amsmath,amsfonts}
\usepackage{amssymb}
\usepackage{enumerate,letltxmacro}
\usepackage{graphicx}
\usepackage{cite}
\usepackage{subfigure}
\usepackage{multirow}
\usepackage{algorithm}
\usepackage{algorithmic}
\usepackage{mathtools}
\usepackage{hyperref}
\usepackage{verbatim}
\usepackage{xcolor}

\hypersetup{draft}

\IEEEoverridecommandlockouts

\newtheorem{theorem}{Theorem}
\newtheorem{definition}{Definition}
\newtheorem{lemma}{Lemma}
\newtheorem{corollary}{Corollary}
\newtheorem{proposition}{Proposition}

\newtheorem{remark}{Remark}

\newcommand{\set}[1]{\mathcal{#1}}
\newcommand{\defined}{\triangleq}
\newcommand{\Real}{{\mathbb{R}}}

\newcommand{\nodes}{\set{V}}

\newcommand{\N}{\set{V}}

\newcommand{\X}{\set{X}}

\newcommand{\A}{\set{A}}
\newcommand{\B}{\set{B}}
\newcommand{\C}{\set{C}}
\newcommand{\D}{\set{D}}

\newcommand{\Y}{\set{Y}}

\newcommand{\G}{\mathbf{G}}
\newcommand{\h}{{h}}
\newcommand{\zero}{\mathbf{0}}

\def\K{{\mathcal K}}

\def\V{{\mathcal V}}

\title{A Minimal Set of Shannon-type Inequalities for Functional Dependence Structures}
\author{\IEEEauthorblockN{Satyajit Thakor$^{\dagger}$,~Terence Chan$^{\ddagger}$ and Alex Grant$^{*}$}
\IEEEauthorblockN{School of Computing and Electrical Engineering, Indian Institute of Technology Mandi$^{\dagger}$\\
Institute for Telecommunications Research, University of South Australia$^{\ddagger}$ \\
Myriota Pty Ltd$^{*}$}
}

\begin{document}

\maketitle
\begin{abstract}
 The minimal set of Shannon-type inequalities (referred to as elemental inequalities), plays a central role in determining whether a given inequality is Shannon-type. Often, there arises a situation where one needs to check whether a given inequality is a constrained Shannon-type inequality. Another important application of elemental inequalities is to formulate and compute the Shannon outer bound for multi-source multi-sink network coding capacity. Under this formulation, it is the region of feasible source rates subject to the elemental inequalities and network coding constraints that is of interest. Hence it is of fundamental interest to identify the redundancies induced amongst elemental inequalities when given a set of functional dependence constraints. In this paper, we characterize a minimal set of Shannon-type inequalities when functional dependence constraints are present.
\end{abstract}

%\begin{IEEEkeywords}
%Network coding capacity bounds, information inequalities, linear programming optimization.
%\end{IEEEkeywords}
\section{Introduction}

Shannon-type inequalities (also called \emph{basic} inequalities) are critical tools to  obtain converse coding theorems (or outer bounds) for the capacity of communication systems. Often the structure of a communication network induces additional functional dependence constraints on the random variables involved in the system model. In \cite{Yeu97}, Yeung gave a framework for information inequalities and characterized a minimal set of Shannon-type inequalities for random variables (in the absence of further functional dependency structures). The characterization of these inequalities provides a mechanical framework for proof of information inequalities and numerical computation of outer bounds for communication networks.  
A notable example is the explicitly computable outer bound (often called the Linear Programming bound or LP bound) for the multi-source multi-sink network coding problem (see \cite{Yeu08} and \cite{YeuZha99}). 
A related problem is to determine whether a given information inequality is Shannon-type (i.e. is implied by the Shannon inequalities). This is a redundancy check problem. A computer program called Information Theoretic Inequality Prover (ITIP) \cite{YeuYanITIP} is available to solve such linear programs. 

In this context, the problems of (a) computing the network coding LP bound, and (b) proving basic information inequalities can both be formulated as linear optimizations with the elemental inequalities as a subset of the constraints. One practical challenge in solving these optimization problems is the large number of variables and constraints. The number of elemental inequalities grows exponentially with the number of variables, making it challenging and sometimes computationally infeasible to generate the set of constraints directly. Therefore, it is of fundamental importance to reduce complexity by eliminating redundant constraints.  In this paper, we address this problem by characterizing a minimal set of Shannon-type inequalities subject to the presence of additional functional dependence constraints. 

Section \ref{sec:background} presents the framework for information inequalities, elemental inequalities, set-theoretic interpretation of information measures and some applications of the set of elemental inequalities under equality constraints such as functional dependence. In Section \ref{sec:MainResults}, we characterize a set of redundant elemental inequalities for a given functional dependence structure. Theorem 1 provides characterization of minimal elemental inequalities for functional dependence constraints and the proof is presented in Section \ref{sec:ThmProof}. In Section \ref{sec: Applications and Future Work}, we discuss some applications of the main results and future directions.

\section{Background}\label{sec:background}

\subsection{Entropy space}

For a set of random variables $\nodes=\{A,B,\dots\}$ with $|\nodes|=n$, let $h:2^{\V}\mapsto\Real$ be a real-valued function  defined on the non-empty subsets of $\nodes$\footnote{With a slight abuse of notation, $2^{\A}$ denotes the set of all non-empty subsets of $\A$ rather than the power set of $\A$. Singletons are represented without braces.}. 
The function $h$  can also be viewed as a point in a $2^n-1$ dimensional Euclidean space, where the  non-empty subsets of $\nodes$ are the indexes of the coordinate axes. This space is the \emph{entropy space} $\mathcal H$~\cite{Yeu08}.  
For notational simplicity, represent $h$ as  a vector
\begin{equation*}
  h =[h(\A): \A \subseteq \nodes \setminus \emptyset]^{\textsf{T}} \in \Real^{2^n-1}.
\end{equation*}
Let $\Gamma$ be the set of all vectors satisfying the elemental Shannon-type inequalities. These \emph{basic} inequalities define the polymatroid axioms:
\begin{align}
 H_{h}(A\mid\V \setminus A) &\geq 0, \quad A \in \V \label{eq:minimal_h} \\
  I_{h}(A;B\mid\C) &\geq 0, \quad A \neq B,  \C \subseteq \nodes \setminus \{A,B\} \label{eq:minimal_I}
\end{align}
where 
\begin{align*}
H_{h}(A | \B) &\defined h(A \cup \B) - h(\B) \\
I_{h}(A;B|\C) &\defined h(A\cup\C) + h(B\cup\C) - h({{A\cup B\cup\C}}) - h({{\C}}).
\end{align*} 

In cases when the vector $h$ is understood implicitly, we will denote 
$H_{h}(A | \B)$ and $I_{h}(A;B|\C)$ simply as $H(A | \B)$ and $I(A;B|\C)$.

The region $\Gamma$ is a polyhedron. In particular it is a pointed cone in the non-negative orthant $\Real_{+}^{2^{n}-1}$. 
We refer to \eqref{eq:minimal_h} resp. \eqref{eq:minimal_I} as the
non-decreasing resp.  submodular \emph{elemental} elemental inequalities. Straightforward enumeration shows that there are
\begin{equation}\label{eq: Gamma ineq}
  m = n + \binom{n}{2}2^{n-2}
\end{equation}
elemental inequalities. It has been proved that these elemental inequalities are non-redundant and that every basic inequality is implied by this set \cite{Yeu91}. 

As the inequalities \eqref{eq:minimal_h}--\eqref{eq:minimal_I} are linear,  the set  $\Gamma$ can be written
in matrix form as
\begin{align}\label{eq:GammaMatrixForm}
  \Gamma \triangleq \{\h: \G\h \geq \zero \}
\end{align}
where $\G$ is a $m\times 2^{n}$ matrix with entries from
$\{-1,0,1\}$, and $\mathbf{0}$ is a length $2^n $ all-zero
vector. Each row of $\G$ encodes one elemental inequality. The
ordering of columns in $\G$ is consistent with the coordinates of $\h$,
e.g., lexicographical ordering on subsets of $\nodes$.

%%%%%%%%%%%%%%%%%%%%%%%%%%%%%%%%%%%%%%%%%%%%
\subsection{Entropy characterization using atoms}

An alternative geometric representation based on a set-theoretic interpretation of information measures was given in \cite{Yeu91}, which we will re-state below. 

For each variable $A$ in $\V$, it corresponds to a set labelled as $\tilde{A}$. Similarly, for a subset of variables $\C$ of $\V$, we will use $\tilde{\C}$ to denote the corresponding union of all sets where $A \in \C$. In other words,
\[
\tilde{\C} = \bigcup_{A\in\C} \tilde{A}.
\]

For a given function $h$, it is associated with a signed measure $\mu_h$ (or $\mu$ for short) such that  for any $\C \subseteq \V$, 
\[
h(\C) = \mu(\tilde{\C}) .
\] 
Here, $\mu(\tilde{\C})$ is the signed measure for the set $\tilde\C$.

An atom is a set of the following form
\begin{align}\label{eq5}
\bigcap_{ {A}\not\in  {\C}} \tilde{A} \setminus \tilde{\C}
\end{align}
where  $\C$ is a proper subset of $\V$. To simplify notation, will denote the atom defined in \eqref{eq5} as $[\C]$.

There are in total $2^{n}-1$ atoms. 
It has been proved in \cite{Yeu91} that the signed measure for the atoms is uniquely determined from $h$ (and vice versa). In addition, there is also a one-to-one correspondence between Shannon's information measures and a unique signed measure denoted $\mu$.  Following the convention in \cite{Yeu91}, 
\begin{align*}
H_{h}(\A|\C) & = \mu(\tilde{A}\setminus \tilde\C)\\
I_{h}(\A ; \B|\C) & = \mu( \tilde{\A} \cap  \tilde{\B} \setminus \tilde\C  ).
\end{align*}

Further, define 
\begin{align}
T_{h}(\alpha)  \triangleq \mu([\alpha]), \alpha\subsetneq\V.
\end{align}
In other words, $T_{h}(\alpha)$ is the signed measure for the atom $[\alpha]$ induced by $\mu$ (or accordingly by the function $h$).  It is easy to see that  
\begin{align}\label{eq43}
h(\beta) = \sum_{\alpha: \beta \setminus \alpha \neq \emptyset} T_{h}(\alpha).
\end{align}

%%%%%%%%%%%%%%%%%%%%%%%%%%%%%%%%%%%%
\subsection{Optimization under functional dependencies}

\begin{definition}[Functional dependency]
A \emph{functional dependency} is a binary 
tuple  $(\X, \Y)$ where $\X, \Y $ are disjoint subsets of $\N$.
It means that the set of variables indexed by $\X$ are functionally imply those by $\Y$.

Further, a polymatroid $h$ (satisfying the basic inequalities) satisfies the functional dependency $(\X, \Y)$ if and only if 
\begin{equation}\label{eq:fdpoly}
H_{h}(\X | \Y) = 0.
\end{equation}
\end{definition}

Let $\Phi$ be a set of $L$ functional dependencies
\begin{equation}\label{eq:fddef}
\Phi \triangleq \{ (\X_\ell, \Y_\ell) \}_{\ell=1}^L
\end{equation}

Consider  the following optimization problem
\begin{multline}\label{eq:optprob}
\text{Optimize } f(\h) \text{ subject to} \\
 \begin{cases}
 h \in \Gamma \\
 h \text{ satisfies functional dependencies in } \Phi. 
 \end{cases}
\end{multline}
Note that $h\in\Gamma$ constrains $h$ to be a polymatroid and the test~\eqref{eq:fdpoly} applies. 

In the context of proving Shannon-type constrained inequalities, $\Phi$ can be given a set of functional dependency constraints. In a network coding problem, $\Phi$ is the set of functional dependency constraints induced by the network topology and the multicast requirement.  

Since all of these constraints are linear inequalities or equalities,~\eqref{eq:optprob} is a linear optimization problem if the objective function $f$ is also linear.

Proving whether a given inequality is basic is a \emph{redundancy check problem}. %That is to check whether, given a system of elemental inequalities, an inequality is redundant or not. 
More generally, a constrained information inequality $\mathbf{b}^{\textsf{T}}{h}\geq 0$ is redundant  (subject to given functional dependencies $\Phi$)  if the minimum value of the linear program
\begin{multline*}
\text{Minimize } \mathbf{b}^{\textsf{T}}{h} \\
  \text{Subject to} 
\begin{cases}
 h \in \Gamma \\
 h \text{ satisfies functional dependencies in } \Phi. 
 \end{cases}
\end{multline*}
is zero.

In general, the set of constraints in~\eqref{eq:optprob} can have a very large and complex structure. Therefore, it is desirable to simplify, or reduce these constraints prior to numerical solution. In this paper, we take a first step to achieve this goal by exploiting the functional dependency structure induced by $\Phi$.

\begin{remark}
Functional dependence relations naturally define equivalence classes on the set of joint entropies. These equivalence classes can be used as a basis for describing all constraints and thus reduce the dimension of the optimization problem. See Section \ref{sec: Applications and Future Work} for further discussion.
\end{remark}

\newcommand{\cl}[1]{{\text{cl}({#1})}}

%%%%%%%%%%%%%%%%%%%%%%%%%%%%%%%%%%%%
\section{Main results}\label{sec:MainResults}
\begin{definition}[Closure]
Let $\C \subseteq \V$. Its \emph{closure} $\cl{\C}$ subject to a given set of functional dependencies $\Phi$
 is the maximal set $\D$  such that 
\[
H_h(\D | \C) = 0 
\]
for all polymatroids $h$ satisfying $\Phi$.
\end{definition}

For any subset of random variables $\C$, its closure is the largest set of random variables that will be  be functionally implied by $\C$, for every set of random variables satisfies the functional dependencies $\Phi$. 

\begin{definition}[Close Atoms]
An atom $[\A]$ is called \emph{close} with respect to a set of functional dependencies $\Phi=\{ (\X_\ell, \Y_\ell )\}_{\ell=1}^L$, if 
 $X_\ell \subseteq \A$ whenever  $\Y_\ell \subseteq \A$, $\ell=1,2,\dots,L$.
\end{definition}

\begin{definition}[Vanishing atoms]
An atom $[\C]$ is called \emph{vanishing} with respect to the functional dependencies $\Phi$ if 
\[
T_h(\C) = 0
\]
for all polymatroids $h$ satisfying $\Phi$. 
\end{definition}

\begin{proposition}[Vanishing atoms]
An atom $[\A]$ is vanishing subject to functional dependencies $\Phi$ if and only if $[\A]$ is not close subject to $\Phi$.
\end{proposition}

\begin{proposition}
A polymatroid  $h$ satisfies all functional dependencies in $\Phi$ if and only if $T_h(\C) = 0$ for all vanishing (non-close) atoms. 
\end{proposition}

In the absence of functional dependency constraints, the set of minimal Shannon inequalities was obtained in \cite{Yeu91} as
\begin{align*}
H(A | \N \setminus A) & \ge 0 
\end{align*}
and
\begin{align*}
I(A;B | \C)  & \ge 0 
\end{align*}
where $A, B \in \N$ and $\C $ is a subset of $\N$.

Subject to further functional dependency constraints $\Phi$, some of these inequalities may become redundant. For example, it can be proved easily that if $\cl{\C} = \cl{\D}$, then 
\[
I(A;B | \C) = I(A;B | \D)
\]
and hence $I(A;B | \C) \ge 0 \iff I(A;B | \D) \ge 0 $.
In the following, we aim to identify such redundant inequalities. 

\begin{lemma}\label{lemma1}
Let $\Phi$ be a given set of functional dependency constraints. 
If subject to $\Phi$, 
$$\cl{\C} = \cl{\C'}$$ and $$\{ \cl{AC}, \cl{BC} \} = \{ \cl{A' \C'}, \cl{B' \C'} \},$$ 
then 
$I(A;B | \C) = I(A';B' | \C')$.
Consequently, 
\[
I(A;B | \C) \ge 0  
\iff
I(A';B' | \C') \ge 0.
\]
\end{lemma}

Lemma \ref{lemma1} illustrates that two inequalities, distinct in the absence of functional dependencies, can become equivalent when functional dependencies are introduced. This paper will identify all such redundant inequalities. 
 
\begin{definition}[Equivalence]
Let $\C$ be close with respect to a given list of functional dependencies $\Phi$ and $A, B \not\in \C$. If 
$B \in \cl{A\C}$ and $A \in \cl{B\C}$, then we say
$A \sim_{\C} B $. 
\end{definition}

It is easy to see that $\sim_{\C}$ is an equivalence relation on $\N \setminus \C$. The relation $A\sim_{\C} B $ means that $A $ and $B$ imply each other when conditioning on $\C$.

\begin{definition}[Minimal atom]
Let $\C $ be close with respect to a given list of functional dependencies $\Phi$. 
A variable $A \in \N \setminus \C$ is called \emph{$\C$-minimal} if whenever there exists $B \in \N \setminus \C$ such that  $B \in \cl{A\C}$, then $A \in \cl{B\C}$.

\end{definition}

\begin{proposition}[Reduction 1]
Consider the inequality
\begin{align}\label{eq11}
I(A;B | \C) \ge 0.
\end{align}
If $A$ is not $\C$-minimal with respect to functional dependencies $\Phi$, then~\eqref{eq11} is redundant.
\end{proposition}
%%%
\begin{IEEEproof}
If $A$ is not $\C$-minimal, then by definition there exists $D \not \sim_{\C} A$ such that 
\[
D \in \cl{A \C}.
\]
In other words, 
\[
H_h(D| A\C) = 0
\]
for all polymatroids $h$ satisfying $\Phi$. Consequently, 
\[
I_h(A;B | \C) = I_h(D;B | \C) + I_h(A ; B |\C, D ).
\]
Thus, \eqref{eq11} is implied by the inequalities
\[
I(D;B | \C) \ge 0
\]
and
\[
I(A ; B |\C, D ) \ge 0.
\]
\end{IEEEproof}

\begin{corollary}
Similarly, the inequality 
\begin{align}\label{eq12}
H(A  | \C) \ge 0
\end{align}
is redundant subject to functional dependencies $\Phi$ if $A $ is not $\C$-minimal with respect to $\Phi$.
\end{corollary}
 
\begin{proposition}[Reduction 2]
Let $A$ and $B$ be $\C$-minimal with respect to given functional dependencies $\Phi$, and suppose $A \sim_{\C} B$.
Then
\begin{align}\label{eq:red2}
I(A; B | \C) \ge 0
\end{align}
is redundant if there exists $\C$-minimal $D$
such that $D \not\sim_{\C} A$. 
\end{proposition}
% PROOF 
\begin{IEEEproof}
\begin{align}
I(A; B | \C) & = H(A  | \C) \\
& = H(A  | \C , D )  +  I(A; D | \C)  
\end{align}
The inequality  $H(A| \C ,D ) \ge 0 $
is implied by $I(A ; B  | \C , D ) \ge 0$ or more precisely 
$I(A ; B  | \cl{ \C D} ) \ge 0$.
\end{IEEEproof}

The remaining inequalities of interest are of the form
\[
I(A; B | \C) \ge 0
\]
such that $\C$ is close, and $A$, $B$ are both $\C$-minimal.

\begin{proposition}[Reduction 3]
Let $A$ be $\C$ minimal with respect to given functional dependencies $\Phi$. If there exists $\C$-minimal $B \not\in \C$ such that $A \not\sim_{\C} B$ then
\begin{align}\label{eq16}
H(A | \C ) \ge 0 
\end{align}
is redundant. 
\end{proposition}
\begin{IEEEproof}
Notice that 
\begin{align}
H(A | \C ) = H(A | B\C)  + I(A;B | \C).
\end{align}
Hence, \eqref{eq16} is implied by 
$H(A | B\C)  \ge 0$ and $I(A;B | \C) \ge 0$.
\end{IEEEproof}

In the above propositions, we have identified numerous redundant inequalities. The following theorem summarises above results by charactersing a minimal set of inequalities that characterise all polymatroids satisfying the functional dependencies $\Phi$.

\begin{theorem}[Minimal characterization]\label{thm}
A function $h$ is polymatroidal and satisfies all functional dependencies in $\Phi$ if and only if it satisfies every Type 1 and Type 2 inequality below:

Type 1:
\begin{align} 
I(A;B | \C) &\ge 0  
\end{align}
where $\C$ is close, $A\not\sim_{\C}B$ and $A,B$ are $\C$-minimal.

Type 2:
\begin{align} 
H(A | \C) &\ge 0
\end{align}
where $\C$ is close and $A$ is $\C$-minimal such that $B \sim_\C A$ whenever $B$ is also $\C$-minimal.

Moreover, this set of inequalities are minimal, in the sense that each Type 1 and Type 2 inequality is non-redundant.
\end{theorem}
 
Again, we take the convention that two inequalities 
\[
I(A;B | \C) \ge 0 
\]
and 
\[
I(A';B' | \C') \ge 0 
\]
are equivalent, if 1) $\cl{\C} = \cl{\C'}$, and 2) 
either $A \sim_{\C} A' $ and  $B \sim_{\C} B' $, 
or $A \sim_{\C} B' $ and  $B \sim_{\C} A' $.

Similarly for Type 2, inequalities  
\[
H(A | \C) \ge 0 
\]
and 
\[
H(A'  | \C') \ge 0 
\]
are deemed equivalent, if 1) $\cl{\C} = \cl{\C'}$, and 2) 
$A \sim_{\C} A' $.

%%%%%%%%%%%%%%%%%%%%%%%%%%%
\section{Proof of Theorem \ref{thm}}\label{sec:ThmProof}

Since the set of inequalities  in Theorem \ref{thm} is obtained by eliminating all redundant inequalities, they certainly will still characterise all polymatroids satisfying the functional dependencies $\Phi$. In the following, we will prove that our obtained Type 1 and Type 2 inequalities are indeed minimal. 

To prove the theorem, we will show that for each Type 1 or Type 2 inequality, we can construct a function $h$ that 1) violates  the chosen inequality,   2) satisfies all remaining Type 1 and Type 2 inequalities, and 3) satisfies all functional dependencies.

\subsection{Type 1 inequalities}
 
Consider a Type 1 inequality of the form 
\begin{align}\label{thm1:eq:type1}
I(A;B | \C) \ge 0.
\end{align}
By definition, 1) $\C$ is close, 2) $A$ and $B$ are $\C$-minimal, and 3)  
$A \not\sim_{\C} B$.

To prove that the above Type 1 inequality is non-redundant, we will construct a 
 function $h$ satisfying all the functional dependencies and all the polymatroidal inequalities 
except \eqref{thm1:eq:type1}. Instead of directly defining $h$, we define its corresponding ``atomic'' function $T_{h}$ as follows:
\begin{align}\label{eq42}
T_{h} ( \beta ) = 
\begin{cases}
-1 & \text{ if } \beta = \C \\
0 & \text{ if } A,B \not\in \alpha, \: \alpha \neq \emptyset, \text{ and } \beta = \C \cup \alpha \\ 
0 & \text{ if } \beta  \text{ is vanishing} \\
2 & \text{ otherwise. } 
\end{cases}
\end{align}
Note that, by definition, 
\begin{align}%\label{eq43}
h(\beta) = \sum_{\alpha: \beta \setminus \alpha \neq \emptyset} T_{h}(\alpha).
\end{align}  
Now, we will show that \eqref{thm1:eq:type1} is indeed non-redundant.

First, we will show that the so constructed function $h$ (or equivalently its atomic version $T_{h}$) violates the inequality \eqref{thm1:eq:type1}.
Note that
\[
I_{h}(A;B | \C) = \sum_{\alpha \supseteq \C : A,B  \not\in\alpha  }T_{h}(\alpha).
\]
It can be verified directly that $I_{h}(A;B | \C) = -1 \le 0$, and hence violating \eqref{thm1:eq:type1}. Next, we will prove that function $h$  satisfies  all other Types 1 and 2 inequalities, and also the functional dependencies. 

From \eqref{eq42}, $T_{h} ( \beta ) = 0$ if $\beta$ is vanishing. Hence, the function $h$ satisfies all the functional dependencies. Now, let us consider a Type 1 inequality 
\begin{align}\label{eq44}
I(i, j | \K) \ge 0
\end{align}
which is different from \eqref{thm1:eq:type1}.
Again, by definition, 1) $\K$ is close,  2) $i$ and $j$ are $\K$-minimal, and 3)  
$i \not\sim_{\K} j$. Notice that  
\[
I_{h}(i;j | \K) = \sum_{\alpha \supseteq \K : i,j  \not\in\alpha  }T_{h}(\alpha).
\]

We will prove that $h$ satisfies \eqref{eq44} by considering different cases. In the first case,  $\C$ does not contain $\K$ as a subset. In this case,  the inequality  \eqref{eq44} will not involve the atom $\C$, and hence will be satisfied by $h$. 

Now, suppose that  $\K$ is a proper subset of $\C$. In this second case, $\K \neq \C$. As $\K$ is close, it is non-vanishing. Therefore, $T_{h}(\K) =2$. Thus, $I_{h}(i;j | \K)  \ge 1$ and hence \eqref{eq44} is satisfied by $h$. It now remains to consider the third case when
$\K=\C$.

If \eqref{eq44} is different from \eqref{thm1:eq:type1},  then 
$$
\{ \cl{i\C}, \cl{j\C} \} \neq  \{ \cl{A\C}, \cl{B\C} \}.
$$ 
In addition, as $A \not\sim_{\C} B$ and $i \not\sim_{\C} j$,   we may also assume without loss of generality that   $ \cl{i\C} \neq \cl{A\C} $ and $ \cl{j\C} \neq \cl{A\C} $. 
In that case,   $\cl{A \C} \supseteq \C$ and $i, j  \not\in  \cl{A \C}$.  Therefore, the non-vanishing atom $\cl{A \C} $ is involved in the inequality \eqref{eq44} and hence the inequality is satisfied by $h$. 

So far, we have proved that $h$ satisfies all Type 1 inequalities except \eqref{thm1:eq:type1}. Now, it remains to show that $h$ also satisfies all Type 2 inequalities. 

Consider a Type 2 inequality 
\begin{align}\label{eq45}
H( i | \K ) \ge 0.
\end{align}
By definition,  $\K$ must be close such that $i$ is $\K$-minimal, and $i\sim_\K j$ whenever $j$ is also $\K$-minimal.
Again, if $\K$ is not a  subset of $\C$, then the inequality does not involve the atom $\C$. Hence, $H_{h} ( i | \K ) $ is nonnegative. On the other hand, since both $A$ and $B$ are $\C$-minimal and $A\sim_\C B$, $\K$ is not equal to $\C$.   So, it remains to consider the case where $\K$ is a proper subset of $\C$. By definition, $T_{h}(\K) = 2$. Hence, it is obvious that from the definition that 
$H_{h}( i | \K ) \ge 1$. The function $h$ thus satisfies all Type 2 inequalities. And the non-redundancy of Type 1 inequalities have been proved.

%%%%%%%%%%%%%%%%%%%%%%%%%%%%%%% 
\subsection{Type 2 inequality}
Next, we will prove that Type 2 inequalities are also non-redundant.
Consider a Type 2 inequality of the form 
\begin{align}\label{eq47}
H( A | \C ) \ge 0.
\end{align}
By definition,  1) $\C$ must be close   and  2) $A$ is $\C$-minimal such that    $A\sim_\C B$ whenever $B$ is also $\C$-minimal.

For this inequality, we define $h$ as follows: 
\begin{align}\label{eq48}
T_{h} ( \beta ) = 
\begin{cases}
-1 & \text{ if } \beta = \C \\
0 & \text{ if } A \not\in \alpha, \: \alpha \neq \emptyset, \text{ and } \beta = \C \cup \alpha \\ 
0 & \text{ if } \beta \text{ is vanishing} \\
2 & \text{ otherwise. } 
\end{cases}
\end{align}
Again, $h$ can be directly obtained via   \eqref{eq43}.

Recall that 
\begin{align} 
H_{h}( A | \C )  = \sum_{\alpha \supseteq \C :  A \not\in\alpha}  T_{h}(\alpha).
\end{align}
From our construction, it is not difficult to see that $H_{h}( A | \C )  = -1$ and hence $h$ does not satisfy the inequality \eqref{eq47} but satisfies all functional dependencies.  Now, it remains to prove that $h$ satisfies all other Type 1 and Type 2 inequalities.

First, consider a Type 1 inequality 
\begin{align}
I(i, j | \K) \ge 0.
\end{align}

If $\K$ is not a subset of $\C$, then  the inequality does not involve the atom $[\C]$. Hence, $I_{h}(i, j | \K) $ is nonnegative. 
Also, as $\C \neq \K$, $T_h(\K)  = 2 $ and hence  $I_h(i, j | \K) \ge 0$.
Now, suppose $\K$ is a proper subset of $\C$.
By definition, $T_{h}(\K) = 2$. Hence,  
$I_h(i, j | \K) \ge 0$. The function $h$ thus satisfies all Type 1 inequalities.

Next, we consider a Type 2 inequality 
\begin{align}\label{eq51}
H( i | \K) \ge 0.
\end{align}
If $\K = \C $, then $i \sim_\K A$ and thus 
\eqref{eq51} and \eqref{eq47} are the same inequality. 
Suppose  $\K \neq \C$. If $\K$ is a proper subset of $\C$, then   $T_h(\K) = 2$. And if $\K$ is not a subset of $\C$, then $T_h(\C)$ is not involved in the inequality \eqref{eq51}. In any cases, this implies that  $h$ will satisfy \eqref{eq51}.
Non-redundancy of \eqref{eq47} and also Theorem \ref{thm} is thus proved.

\section{Applications and Future work}\label{sec: Applications and Future Work}
%The complete functional dependency structure can be obtained by checking for functional dependencies given $\Gamma \cap \Delta$ using the linear programming framework \cite{Yeu97}. When the number of random variables is large, such an approach is not suitable since the size (both the rows and the columns) of the matrix representing elemental inequalities increases exponentially. 
It is desirable to obtain directly the reduced matrix representing the %region $\Gamma \cap \Phi$ for given functional dependency constraints. 
functions $h$ in the constraint region of \eqref{eq:optprob}.
In \cite{ThaGraCha09} (see also \cite{ThaGraCha16a}, \cite{Tha12}), we gave a graph based recursive algorithms to find implied functional dependencies from local functional dependencies. Though the graph based algorithm does not always give all implied functional dependence relations, it gives many functional dependencies depending on the structure of local functional dependencies without depending on the linear programming framework. In \cite{ThaGraCha11} (see also \cite{Tha12}), we gave algorithms to directly obtain a reduced size matrix defining the constraint region.
Despite the fact that the reduction was not minimal, it was demonstrated that for the well known butterfly network, the matrix size can be reduced by $98\%$. 
The number of variables, $2^n-1$, in the optimization problems can be reduced to the number of equivalence classes for a given functional dependency structure. Given the functional dependence structure, the\textit{ minimal} set of inequalities defining %$\Gamma \cap \Phi$
the constraint region can also be obtained in a matrix form directly using the approach similar to  \cite{ThaGraCha11}. Applications of this matrix include solving the optimization problem \eqref{eq:optprob}.

As a continuation of research work in this direction, we aim to employ the results of this paper to develop more refined algorithms (compared to \cite{ThaGraCha11}), for obtaining the ``minimal'' matrix directly. Moreover, we are investigating further generalizations of the ideas presented in this paper.
%\section{Conclusion}

\section*{Acknowledgment}
This work is supported in part by Science and Engineering Research Board, Department of Science and Technology, Government of India, under project SB/S3/EECE/265/2016. It is also supported in part by Australian Research Council under Discovery Project DP150103658. 
%\end{IEEEproof}

\bibliographystyle{ieeetr}
\bibliography{network}

\end{document}